\documentclass[aps,pre,twocolumn,floatfix]{revtex4-2}

\usepackage{amsmath}
\usepackage{amssymb}
\usepackage{graphicx}
\usepackage{dcolumn}
\usepackage{bm}
\usepackage[usenames,dvipsnames]{xcolor}
\usepackage{multirow}
\usepackage{hyperref}
\usepackage{epstopdf}

\usepackage[utf8]{inputenc}
\usepackage[T1]{fontenc}
\usepackage{microtype}
\usepackage{txfonts}
\usepackage{ulem}
\usepackage{tabularx}

\usepackage{url,hyperref}
\hypersetup{colorlinks=true, linkcolor=blue, urlcolor=blue, citecolor=blue}

\def\av#1{\left\langle{#1}\right\rangle}
\def\d{\mathrm{d}}

\def\ol#1{\overline{#1}}

\begin{document}


\title{Charge symmetry broken complex coacervation}

\author{Arghya Majee$^{1,2}$, Markus Bier$^{1,2,3}$, Ralf Blossey$^4$, Rudolf Podgornik$^{5,6}$}
\affiliation{$^1$Max Planck Institute for Intelligent Systems, Stuttgart, Germany\\ 
             $^2 $IV. Institute for Theoretical Physics, University of Stuttgart, Germany\\
             $^3$Fakult\"at Angewandte Natur- und Geisteswissenschaften, Hochschule f\"ur Angewandte Wissenschaften W\"urzburg-Schweinfurt, Germany\\
             $^4$ Universit\'e de Lille, CNRS, UMR8576 Unit\'e de Glycobiologie Structurale et Fonctionnelle (UGSF), Lille, France\\
             $^5$School of Physical Sciences and Kavli Institute for Theoretical Sciences, University of Chinese Academy of Sciences, Beijing, China\\
             $^6$CAS Key Laboratory of Soft Matter Physics, Institute of Physics, Chinese Academy of Sciences, Beijing, China}

\date{December 24, 2020}

\begin{abstract}
Liquid-liquid phase separation has emerged 
as one of the important paradigms in the chemical 
physics as well as biophysics of charged macromolecular 
systems. We elucidate an equilibrium phase separation 
mechanism based on charge regulation, i.e., 
protonation-deprotonation equilibria controlled by pH, 
in an idealized macroion system which can serve as 
a proxy for simple coacervation. {First, a} low-density 
density-functional calculation reveals the dominance 
of two-particle configurations coupled by ion adsorption 
on neighboring macroions. Then a binary cell 
model, solved on the Debye-H\"uckel as well as the 
full nonlinear Poisson-Boltzmann level, unveils the 
charge-symmetry breaking as inducing the phase 
separation between low- and high-density phases 
as a function of pH. These results can be identified
as a \textsl{charge symmetry broken complex coacervation} 
between chemically identical macroions.
\end{abstract}

\maketitle


\section{Introduction\label{sec:1}}

The importance of \textsl{complex coacervation} 
in polymers, colloids and particularly proteins that 
exhibit an associative liquid-liquid phase 
separation (LLPS), driven by electrostatic 
interactions between oppositely charged macroions, 
has been recognized for about a century 
\cite{Tie11, Jon29}, though its fundamental role in 
compartmentalization and intracellular phase 
transitions in biological systems has been identified 
only recently \cite{Bra15}. The electrostatically driven 
attractions, as already hypothesized in the early 
Overbeek-Voorn theory  \cite{Sin20}, and later 
developed within more sophisticated theoretical 
frameworks \cite{Zha18}, have been shown to 
result in LLPS, thus  being recognized as the defining 
feature of complex coacervation \cite{Per19}. On the 
other hand, for like-charged macroions with monovalent 
counterions, it is the variation of the solvent conditions, 
such as temperature, pH and ionic strength \cite{Che20}, 
that modifies the electrostatic repulsion which would 
otherwise prevent coacervation, except when 
countered by the like-charge attraction mediated 
by multivalent counterions \cite{Smi20}. Studies of 
adhesive proteins \cite{Lee11} as well as 
several proteins involved in some protein aggregation 
diseases (e.g., Alzheimer's disease and 
amyotrophic lateral sclerosis) made it clear, 
however, that \textsl{simple coacervation} involving 
only similarly charged macroions can also lead to 
LLPS, presumably because of short range specific 
interactions of non-electrostatic nature \cite{Kim16}. 

The proper understanding of the mechanisms of 
oppositely charged (complex) and similarly 
charged (simple) coacervations --- interesting 
in the context of functional biomimetic and 
adhesive materials of the chemical, pharmaceutical, 
textile and food industries \cite{Sri16}, and 
particularly relevant in the biophysical milieu, 
where different facets of protein chemistry 
\cite{Lun13} can lead to coexisting liquid-like 
states ---  has been claimed to be one of the 
most important problems in the physical 
chemistry of the cytoplasm \cite{Hym14}.

While experimentally well documented, the 
dependence  of the associative LLPS on the 
bathing  environment conditions, such as 
the solution pH \cite{Fra18}, has lacked 
a comprehensive theoretical elucidation 
based on relevant microscopic models. 
That these effects are particularly important 
in protein solutions \cite{Lun13} is clear from the fact 
that the protein charge is not fixed, 
but is a result of the proton-mediated 
dissociation of amino-acid (AA) groups at 
the solvent accessible surface \cite{Bor01}, 
whose chemical equilibrium then depends on 
the bathing environment parameters such as 
the solution pH \cite{Ben90}. The physical 
basis of the protein charging is consequently 
understood as the \textsl{charge regulation 
(CR) mechanism, i.e.}, an association/dissociation 
process that couples the local electrostatic 
potential with the local charge, leading to 
a self-consistent partitioning of the protein 
charge states with pronounced effects also 
on the properties of other macroions such 
as weak polyelectrolyte solution and gel 
conformational as well as charge properties, 
see Ref. \cite{Avn19} for details.

Theoretical analyses of the CR effects in 
the formation of macroion condensates, 
that depend explicitly on the solution pH, 
have been scant. A simple \textsl{cell 
model} approach was used to analyse the 
CR macroions in solution \cite{Gis94}, 
together with their effective charge 
\cite{Boo12}, and the corresponding 
phase behavior \cite{All04}. A thermodynamic 
minimal model analysis was proposed to 
study LLPS in a fixed pH ensemble based 
on a set of reactions describing the 
protonation/deprotonation reactions 
of the solution macroions, conducive 
to multiple charge states \cite{Ara20}. 
The equilibrium charge state and critical 
behavior of CR  macroions was studied 
based on a collective description of a 
solution composed of CR macroions and 
simple salt ions in the bulk \cite{Avn18}. 
Within the mean-field approximation it 
was found that above a critical 
concentration of salt, a non-trivial 
distribution of coexisting charge states 
leads to a liquid-liquid phase separation, 
similar to the behavior of micellar 
solutions close to the critical micelle 
concentration \cite{Avn20}. Extreme 
CR, implying a constant surface potential, 
was also invoked in binary suspensions of 
charged colloids \cite{Eve16}, possibly 
leading to charge-alternating linear strings.

In what follows we will present a detailed 
analysis not only of the liquid-liquid 
transition in CR macroion systems, but 
also the corresponding spatial charge 
distribution that is at its origin. The 
central idea, as depicted in Fig.~\ref{Fig:1}, 
stems from the striking observation \cite{Maj18} 
that a pair of chemically identical interacting 
charge-regulated planar 
macroions are not necessarily equally 
charged and that the electric field at 
the mid-plane of the set-up does not 
necessarily vanish. In order to provide 
a firm basis to the intuitive expectations 
on the charge symmetry-breaking transition 
for a spherical macroion system, we present 
arguments based on {a} \textsl{density 
functional theory} (DFT) as well as on a 
\textsl{binary cell model} (BCM). Moreover, 
we show that the LLPS is based on a 
symmetry breaking transition of the macroion 
charge distribution, characterized by a 
spatially alternating sign of the macroion 
charge. In {that} respect this CR system 
driving a complex coacervation behaves 
not unlike the alternating multilayer 
structure of the electrical double layer 
in ionic liquids \cite{Fed14}, except that 
here the charge alternation is driven by 
CR and not by the presence of different 
ion species. We identify this spatial 
charge layering, stemming from a symmetry 
broken charge distribution and leading to 
phase behavior that exhibits features of 
complex coacervation phenomenology as 
\textsl{charge symmetry broken complex 
coacervation} between chemically identical 
macroions.

\section{Charge regulation model\label{sec:2}}

\begin{figure}[!t]
   \includegraphics[width=8cm]{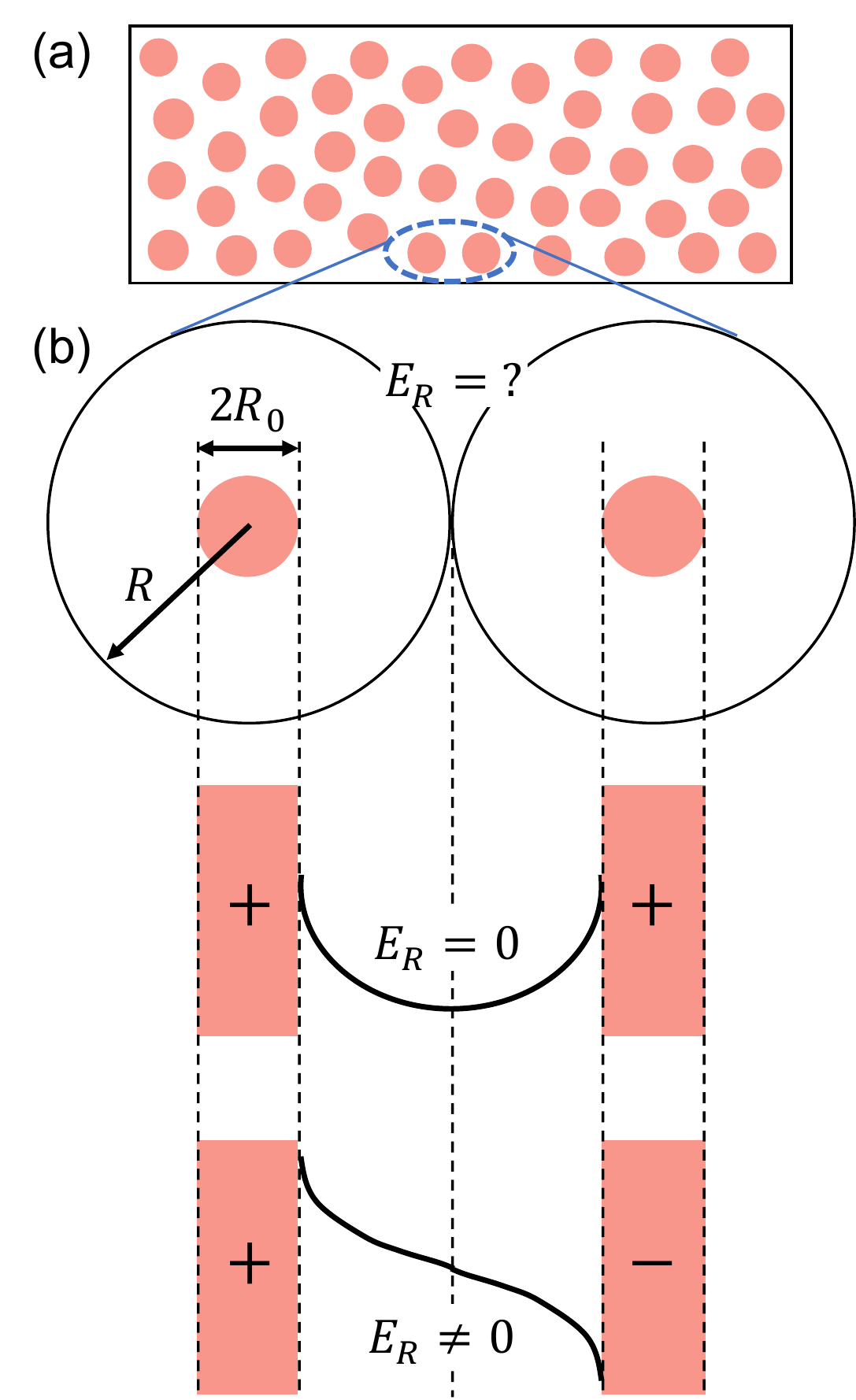}
   \caption{Macromolecular solution (panel (a)) 
   and the magnified view (part (b)) of a small 
   portion of it. Within a {\sl binary cell model}, each 
   macroion (indicated by red circles) of radius 
   $R_0$ is surrounded by a cell of radius $R$ 
   and the interaction between a pair such 
   cell-surrounded spheres is considered. The 
   electric field $E_R$ at the cell boundary is 
   assumed to be uniform and as it is the case 
   for interacting planar surfaces, the value of 
   $E_R$ depends on the charge states of the 
   neighboring macroions forming the pair. The 
   two cells of the binary cell model allow for 
   asymmetric charge configurations ($E_R \neq 0$), 
   which are excluded from the standard symmetric 
   charge (single cell) cell mode ($E_R = 0$). 
   In the latter case the binary cell model then 
   reduces to the standard cell model.}
   \label{Fig:1}
\end{figure}

As shown in Fig.~\ref{Fig:1}(a), consider 
spherical macroions (e.g., proteins, 
polyelectrolytes, colloids etc.) of radius 
$R_0$ - whose surface charge is regulated 
following a mechanism identical to the 
charge-regulation model introduced in 
\cite{Har06, Maj18, Maj19} - that are suspended in 
a univalent salt solution. 
In short, each macroion surface contains a 
fixed number of negative charges and twice 
as many neutral sites where adsorption or 
desorption of protons can take place. 

The surfaces are charge regulated through 
this adsorption-desorption, and the fraction 
$\eta$ of filled sites on a surface is a 
degree of freedom within our model. By 
construction, $\eta\in[0,1]$. If the area 
per site is $a^2$, then the charge density 
is given by 
$\sigma=\frac{e}{a^2}\left(\eta-\frac{1}{2}\right)$ 
with $e>0$ being the elementary charge, so 
that $-{\textstyle\frac12}\frac{e}{a^2}\leq\sigma\leq{\textstyle\frac12}\frac{e}{a^2}$. 
The surface number density $1/a^2$ of adsorption 
sites is related to the number $K=4\pi R_0^2/a^2$ 
of adsorption sites on a single colloidal macroion. 

As in Refs.~\cite{Har06, Maj18, Maj19} 
we base our macroion CR model on the 
Frumkin-Fowler-Guggenheim isotherm \cite{Koo20} 
of the macroion surface defined with the 
phenomenological free energy of a single
adsorption site in the units 
of thermal energy $\beta=1/k_BT$ as
\begin{align}
   \beta\widehat\Omega^{\text{s}}\left(\eta\right) = 
   - \alpha\eta -\frac{\chi}{2}\eta^2 
   + \eta\ln(\eta) 
   + (1-\eta)\ln(1-\eta).
   \label{eq:1}
\end{align}
The parameters $\alpha$ and $\chi$ are 
phenomenological and describe the 
non-electrostatic part of the proton-macroion 
and the proton-proton interactions at the 
macroion surface. In the case of (de)protonation 
reaction, the dependence of $\alpha$ on the 
bulk pH is model specific \cite{Avn18}, 
but one can explicitly identify 
$\alpha=(\mathrm{pK}-\mathrm{pH})\ln 10$, 
where $\mathrm{pH} = -\log_{10}[\mathrm{H}^{+}]$,  
with $[\mathrm{H}^{+}]$ being the proton 
concentration in the bulk and $\mathrm{pK}$ 
is the dissociation constant of the 
deprotonation reaction, in the case of 
the Langmuir adsorption model \cite{Nin71}. 
Furthermore, $\chi$, as in the related lattice 
regular solutions theories (e.g., the 
Flory-Huggins theory \cite{Ter02}) describes 
the  short-range interactions between nearest 
neighbor adsorption sites on the macroion 
surface \cite{Avn20}. An increase in 
the parameter value $\alpha$ encodes a 
favorable adsorption free energy between 
protons and the macroion surface, while 
$\chi\geq0$ represents the tendency of 
protons on the macroion surface adsorption 
sites to phase separate into domains. 

In what follows we will use both $\alpha$ as 
well as $\chi$ as purely phenomenological interaction 
parameters,  quantifying the adsorption energy in 
the surface (de)protonation reactions and the 
nearest-neighbor surface energy of filled surface 
adsorption sites.

The model Eq.~\eqref{eq:1} was applied to 
lamellar-lamellar phase transition in a 
charged surfactant system \cite{Har06} and 
a good correspondence with experiments was 
obtained for the didodecyldimethylammonium 
chloride (DDACl) data with $\alpha = -3.4$, 
$\chi=14.75$, and for the didodecyldimethylammonium 
bromide (DDABr) data with $\alpha =-7.4$ and 
$\chi=14.75$, see Ref. \cite{Har06} for 
details. The same model was successfully 
applied also to other systems, see e.g., 
\cite{Fin17, Fin19}.

\section{Density-functional theory in the low-density limit\label{sec:3}} 

\subsection{Formalism}

As the configuration of a single colloidal macroion 
is described by the position $\mathbf{r}\in\mathcal{V}$ 
of the center of mass and the average degree
of protonation $\eta\in[0,1]$ on its surface, the whole 
suspension can be described by the number density 
$n(\mathbf{r},\eta)$. The equilibrium number density 
minimizes a grand canonical density functional
$\Omega[n]$ (see Ref.~\cite{Eva79}), which is 
approximated in the low-density limit by
\begin{align}
   \beta\Omega[n] =   
   & \int\limits_{\mathcal{V}}d^3r\int\limits_0^1d\eta\,~n(\mathbf{r},\eta)
     \left[\ln\left(\frac{n(\mathbf{r},\eta)}{\zeta}\right) - 1 
     + K\beta\widehat\Omega^{\text{s}}\left(\eta\right)\right]\notag\\
   & +\beta F^\text{ex}_\text{hc}[n] 
     + \beta F^\text{ex}_\text{el}[n].
   \label{eq:2}
\end{align}
Here, $\zeta$ is the fugacity, $F^\text{ex}_\text{hc}$ 
represents the excess free energy due to the hard core 
interaction between two colloidal macroions and 
$F^\text{ex}_\text{el}$ describes the excess free energy
contribution of the electrostatic interaction. In the 
following, the hard core excess free energy 
$F^\text{ex}_\text{hc}$ is based on the Percus-Yevick 
(PY) closure and the corresponding equation of state 
via the compressibility route is used \cite{Han86, McQ00}.

The colloidal macroions are assumed to be suspended 
in an electrolyte solution with relative permittivity 
$\varepsilon_r$ and Debye length $1/\kappa$. For not 
too highly charged macroions in a sufficiently dilute 
suspension one can use the Debye-H\"uckel (DH) 
approximation \cite{Deb23, McQ00} for the electrostatic 
two-particle interaction potential
\begin{align}
   \beta U_\text{el}(r,\eta,\eta')
   = \sigma^*(\eta)\sigma^*(\eta')K^2
   \frac{\ell_B\exp(-\kappa( r-R_0))}
   {\left(1+\kappa R_0\right)r},
   \label{eq:3}
\end{align}
where here and below $r = |\mathbf{r}-\mathbf{r}'|$, 
while the dimensionless surface charge density 
\begin{align}
 \sigma^*(\eta)=\frac{\sigma(\eta)a^2}{e}=\eta-\frac{1}{2}
 \label{eq:4}
\end{align}
of a colloidal macroion with average degree 
of protonation $\eta$ and the Bjerrum length 
$\ell_B = \beta e^2/(4\pi\varepsilon_0\varepsilon_r)$
of the solvent with the vacuum permittivity 
$\varepsilon_0$ are introduced. Considering the 
electrostatic interaction $U_\text{el}$ as a 
perturbation of the hard core interaction 
$U_\text{hc}$ with 
\[
   \beta U_\text{hc}(r) =
   \begin{cases}
      0 & \text{for $r\geq2R_0$} \notag\\
      \infty & \text{for $r<2R_0$}\notag
   \end{cases}
\]
one obtains in the low-density limit \cite{Eva79}
\begin{align}
   \beta F^\text{ex}_\text{el}[n]=
   &\frac{1}{2}\int\limits_\mathcal{V}d^3r \int\limits_\mathcal{V}d^3r'\int\limits_0^1d\eta
   \int\limits_0^1d\eta'n(\mathbf{r},\eta)\,n(\mathbf{r}',\eta')\label{eq:5}
   \\
   &\!\cdot\exp(-\beta U_\text{hc}(\mathbf{r}-\mathbf{r}'))\left(1\!-\exp\left(-\beta
   U_\text{el}(\mathbf{r}-\mathbf{r}',\eta,\eta')\right)\right).\notag
\end{align}

The considered model is then specified 
by the following  five parameters: 
$\alpha, \chi, \kappa R_0, \kappa\ell_B, K$, 
among which $\alpha$ and $\chi$ describe 
the charge regulation (according to 
Sec.~\ref{sec:2}). The values for the 
parameters $\alpha$ and $\chi$ are 
chosen keeping in mind that for 
$\chi=-2\alpha$ the surfaces remain 
charge neutral for $\chi<\chi_c$ below 
a certain critical value $\chi_c>0$, 
whereas they can be oppositely charged 
for $\chi>\chi_c$ \cite{Maj18}. Assuming 
spherical colloidal macroions of radius 
$R_0=10\,\mathrm{nm}$ in an aqueous 
electrolyte solution with ionic strength 
$1\,\mathrm{mM}$, i.e., with Bjerrum length 
$\ell_B\approx0.7\,\mathrm{nm}$ and Debye 
length $1/\kappa\approx10\,\mathrm{nm}$, 
leads to the values $\kappa R_0\approx1$ 
and $\kappa\ell_B\approx0.07$. Finally, 
within the present DFT approach, 
$K\in\{20,40,45,50\}$ adsorption sites 
per colloidal macroion are considered. 
This corresponds to surface areas per 
adsorption site 
$a^2=4\pi R_0^2/K\in\{63,31,28,25\}\,\mathrm{nm^2}$, 
i.e., average distances between 
neighbouring adsorption sites of 
$a\in\{7.9,5.6,5.3,5\}\,\mathrm{nm}$, respectively.


\subsection{Charge-regulation-induced phase separation}

\begin{figure}[!t]
   \includegraphics[width=8cm]{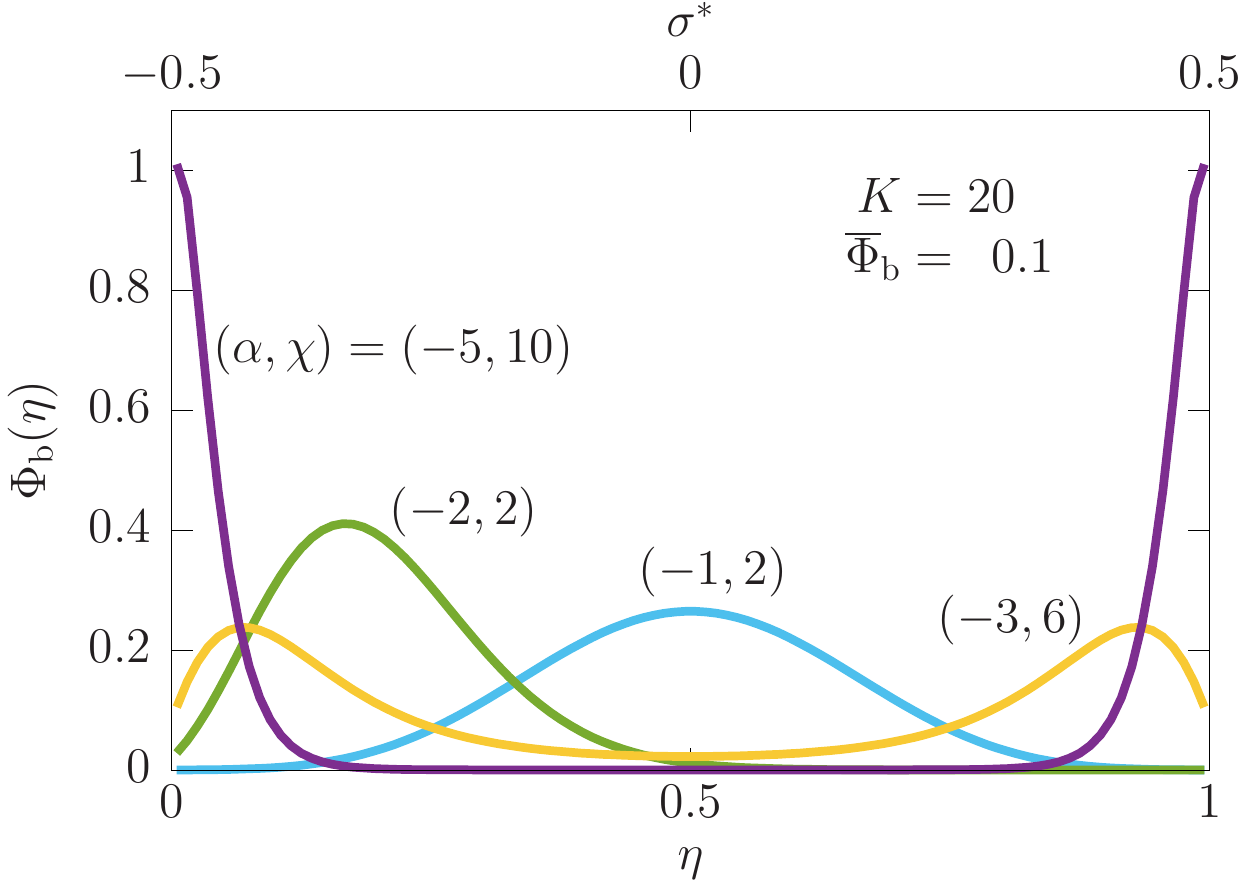}
   \caption{Distribution of surface charges 
   $\sigma^*$ and degrees of protonation 
   $\eta$ as well as the associated bulk 
   packing fraction profile $\Phi_{\text{b}}(\eta)$ 
   (as defined in Eq.~\eqref{eq:6}) in a 
   suspension with packing fraction 
   $\ol{\Phi}_{\text{b}}=0.1$ of colloidal 
   macroions with $K=20$ adsorption sites per 
   macroion. For small values of $\chi>0$ 
   a unimodal distribution is observed, 
   whereas for sufficiently large values 
   of $\chi$ the surface charge distribution 
   becomes bimodal with increasingly large 
   magnitudes of the peak surface charges.}
   \label{Fig:2}
\end{figure}

In the bulk of the colloidal suspension, 
no position dependence occurs for the 
equilibrium density profile as well as 
for the total bulk packing fraction, i.e., 
$n(\mathbf{r},\eta) = n_{\text{b}}(\eta)$ 
and $\ol{\Phi}(\mathbf{r}) = \ol{\Phi}_{\text{b}}$.
Note that uniformity of the bulk profiles 
of a fluid is not an assumption or an 
approximation but the necessary 
consequence of translational invariance. 
Upon solving the Euler-Lagrange 
Equation~\eqref{eq:4_SI} 
given in Appendix \ref{sec:A1} one obtains 
the bulk packing fraction profile
\begin{align}
   \Phi_{\text{b}}(\eta) = \frac{4\pi}{3}R_0^3n_{\text{b}}(\eta),
   \label{eq:6}
\end{align}
which provides the distribution of the 
average degree of protonation $\eta$ or, 
equivalently, of the surface charge 
densities $\sigma^*(\eta)$ (see 
Eq.~\eqref{eq:4}). Figure~\ref{Fig:2} 
displays this distribution for a 
suspension with bulk packing fraction 
$\ol{\Phi}_{\text{b}}=0.1$ of colloidal 
macroions with $K=20$ adsorption sites 
per macroion. For small values of 
$\chi>0$, e.g., $\chi=2$ (see the blue 
and the green curves in Fig.~\ref{Fig:2}), 
the surface charge distribution is 
unimodal, i.e., the colloidal macroions 
are essentially equally charged. If 
the charge regulation parameters 
$\alpha$ and $\chi$ (see Eq.~\eqref{eq:1}) 
fulfill the relation $\chi=-2\alpha$ 
the peak is at $\sigma^*=0$ (see 
the blue curve in Fig.~\ref{Fig:2}), 
whereas for $\alpha\gtrless-\chi/2$ 
the majority of colloidal macroions 
carry a surface charge $\sigma^*\gtrless0$ 
(see the green curve in Fig.~\ref{Fig:2}).
Upon increasing the value of the 
charge-regulation parameter $\chi$ the 
surface charge distribution becomes 
bimodal with the peaks being located 
at increasingly large magnitudes of 
the surface charges (see the yellow 
and the purple curves with the lightest 
and the darkest shades, respectively, in 
Fig.~\ref{Fig:2}). For $\alpha=-\chi/2$ 
both peaks represent the same number 
of macroions, but of opposite charge. 
The presence of equal amounts of 
oppositely charged colloidal macroions 
is expected to lead to compact structures, 
i.e., to a high-density phase.

Recently Avni \textsl{et al}. \cite{Avn20} 
described a two-phase (or even multiple-phase) 
coexistence region(s), where macroions with 
low adsorption site occupation coexist with 
macroions with high site occupation, akin to 
the case presented in Fig.~\ref{Fig:2}. 
However, the model in Ref.~\cite{Avn20} 
differs from the present one in that there 
are two types of adsorption sites, one 
charging positively and one charging 
negatively, on the macroions, whereas 
here the negative surface charges are 
fixed and only the positively charging 
sites are charge regulated.

\begin{figure}[!t]
   \includegraphics[width=8cm]{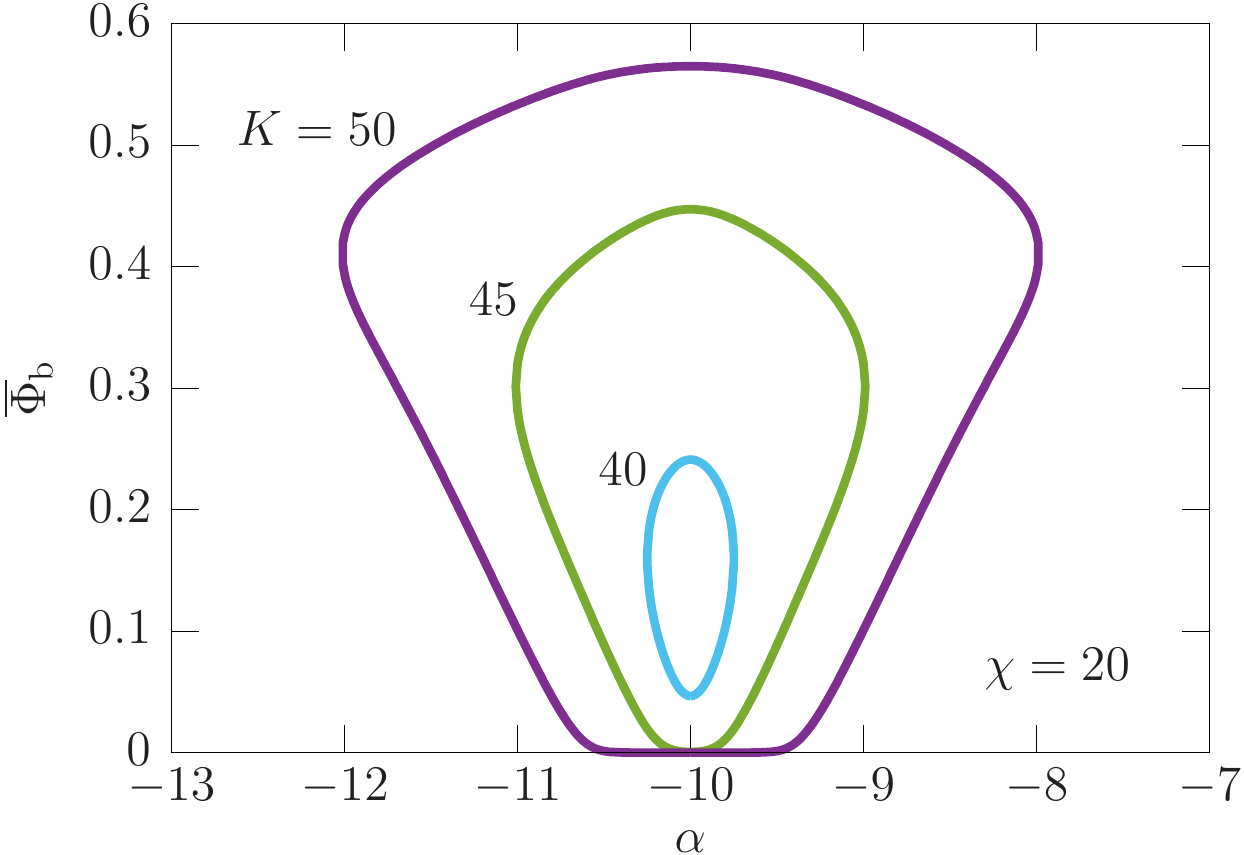}
   \caption{Binodals of the charge-regulation-induced 
   phase separation of colloidal macroions with $K=40$ 
   (blue, light shade), $45$ (green, intermediate 
   shade) and $50$ (purple, dark shade) adsorption 
   sites per macroion. The charge regulation parameter 
   $\chi=20$ is chosen arbitrarily and the 
   pH-sensitive parameter $\alpha\in[-13,-7]$ 
   is tuned around the value $-\chi/2 = -10$, where 
   oppositely charged colloidal macroions are expected 
   to occur. The interior of the loops corresponds 
   to the two-phase regions, where phase separation 
   into a low-density and a high-density phase occurs 
   at the given value of $\alpha$. The two-phase region 
   widens upon increasing the number $K$ of adsorption
   sites per macroion as a result of an increasing 
   magnitude of the electrostatic interaction.}
   \label{Fig:3}
\end{figure}

In order to illustrate the occurrence 
of a phase separation into a high-density 
and a low-density phase for 
$\alpha \approx -\chi/2$, the case 
$\chi=20$ for various numbers $K$ 
of adsorption sites per colloidal 
macroion are considered. Figure~\ref{Fig:3} 
displays the binodals of the 
charge-regulation-induced phase 
separation transition for $K=40$ 
(blue curve, light shade), $45$ 
(green curve, intermediate shade) 
and $50$ (purple curve, dark shade). 
The interior of the loops corresponds 
to the two-phase regions, where phase 
separation into a low-density and 
a high-density phase at the given 
value $\alpha$ occurs. The two-phase 
region widens upon increasing the 
number $K$ of adsorption sites per 
macroion as a result of an increasing 
magnitude of the electrostatic interaction.

It is well-known that there are no 
fluid phases with packing fractions 
above $\ol{\Phi}_{\text{b}}\approx0.5$ 
as then crystallization sets in. This 
phenomenon is not covered within the 
present framework so that values of 
$\ol{\Phi}_{\text{b}}\gtrapprox0.5$ 
here are indicative of colloidal 
aggregation.

The binodals of the charge-regulation 
induced phase separation presented in 
Fig. \ref{Fig:3} are quite similar to 
those calculated by Adame-Arana \textsl{et al}. 
\cite{Ara20}, even if the calculational 
details differ,  the main difference being 
that we include the electrostatic 
interactions explicitly {\sl via} the 
two-body DH interaction, Eq.~\eqref{eq:3}, 
while in Ref.~\cite{Ara20} the charge-charge 
interaction is characterised by a 
phenomenological Flory-like parameter.


\subsection{Fluid structure}

\begin{figure}[!b]
   \includegraphics[width=8cm]{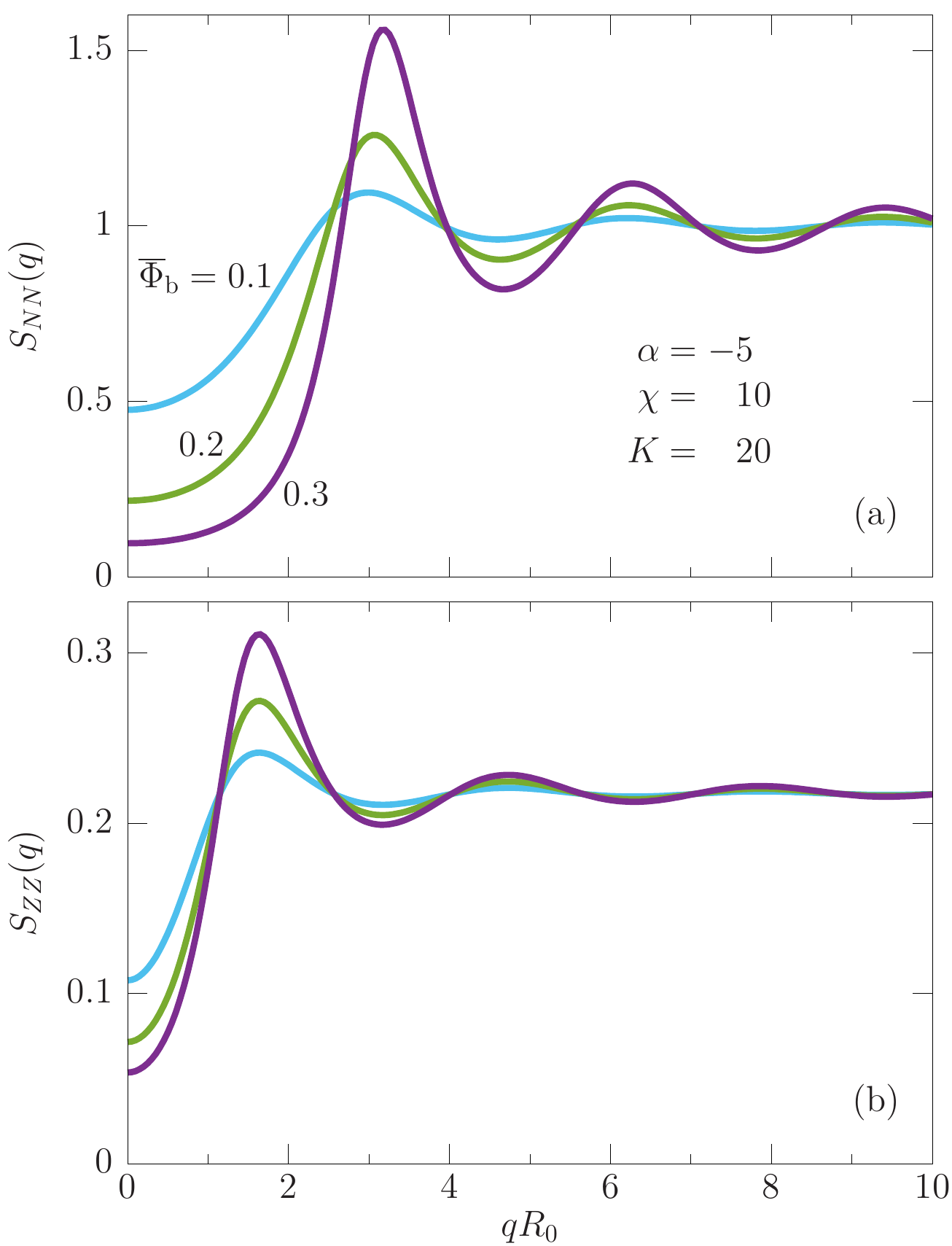}
   \caption{Structure factors $S_{NN}(q)$ 
   (panel (a)) and $S_{ZZ}(q)$ (panel (b)) 
   of suspensions with packing fractions 
   $\ol{\Phi}_{\text{b}} = 0.1$ (blue, light 
   shade), $0.2$ (green, intermediate shade) 
   and $0.3$ (purple, dark shade) of 
   charge-regulated colloidal macroions 
   of radius $R_0$ with $K=20$ adsorption 
   sites per macroion and charge regulation 
   parameters $\alpha=-5$, $\chi=10$. As 
   defined in Eqs.~\eqref{eq:7} and 
   \eqref{eq:8}, the structure factor 
   $S_{NN}(q)$ in panel (a) describes 
   the relative distribution of colloidal 
   macroions irrespective of their charge, 
   whereas $S_{ZZ}(q)$ in panel (b) 
   describes the relative distribution 
   of charge within the fluid (see 
   Table~\ref{Tab:1} for details). The 
   location of the main peak of $S_{ZZ}(q)$ 
   relative to that of $S_{NN}(q)$ indicates 
   an alternating charge structure.}
   \label{Fig:4}
\end{figure}

The bulk structure of the considered 
suspensions of charge-regulated colloidal 
macroions described by the density functional 
in Eq.~\eqref{eq:2} can be inferred from the 
partial structure factor $S(q,\eta,\eta')$ 
(see Appendix \ref{sec:A1} for details).
It can be conveniently analyzed in terms 
of the number-number structure factor
\begin{align}
   S_{NN}(q) = \int\limits_0^1d\eta\int\limits_0^1d\eta'S(q,\eta,\eta'),
   \label{eq:7}
\end{align}
which describes the relative distribution 
of colloidal macroions irrespective of 
their charge, and the charge-charge 
structure factor
\begin{align}
   S_{ZZ}(q) = \int\limits_0^1d\eta\int\limits_0^1d\eta'
   \sigma^*(\eta)\sigma^*(\eta')S(q,\eta,\eta'),
   \label{eq:8}
\end{align}
which describes the relative distribution 
of charge within the fluid.

\begin{table}[!t]
   \begin{tabular}{c||c|c||c|c||c}
   $\ol{\Phi}_{\text{b}}$ &
   $\lambda_{NN}/R_0$ & $\xi_{NN}/R_0$ &
   $\lambda_{ZZ}/R_0$ & $\xi_{ZZ}/R_0$ &
   $N_1$
   \\
   \hline
   \hline
   0.1 & 2.40 & 0.66 & 5.80 & 0.87 & 3.4
   \\
   0.2 & 2.21 & 0.98 & 4.82 & 1.11 & 6.1
   \\
   0.3 & 2.06 & 1.44 & 4.39 & 1.34 & 8.3
   \end{tabular}
   \caption{Characteristics (see main text) 
   of suspensions of packing fractions 
   $\ol{\Phi}_{\text{b}} = \{0.1, 0.2, 0.3\}$ 
   of charge-regulated spherical colloidal 
   macroions of radius $R_0$ with $K=20$ 
   adsorption sites per macroion and charge 
   regulation parameters $\alpha=-5$, 
   $\chi=10$ inferred from the structure 
   factors $S_{NN}(q)$ (see Fig.~\ref{Fig:4}(a)) 
   and $S_{ZZ}(q)$ (see Fig.~\ref{Fig:4}(b)). 
   Upon increasing the packing fraction 
   $\ol{\Phi}_{\text{b}}$ the mean 
   nearest-neighbor distance $\lambda_{NN}$ 
   decreases to almost the
   close-contact distance $2R_0$ of two 
   hard spheres of radius $R_0$, whereas 
   the periodicity $\lambda_{ZZ}$ of the 
   charge distribution is slightly larger 
   than twice that distance, 
   $\lambda_{ZZ}\gtrsim 2\lambda_{NN}$. 
   In parallel the coordination number 
   $N_1$ of colloidal macroions in the 
   nearest-neighbor shell increases.}
   \label{Tab:1}
\end{table}

Figure~\ref{Fig:4} displays the structure 
factors $S_{NN}(q)$ and $S_{ZZ}(q)$ for 
suspensions with packing fractions 
$\ol{\Phi}_{\text{b}}\in\{0.1, 0.2, 0.3\}$ 
of colloidal macroions with $K=20$ 
adsorption sites per macroion and charge 
regulation parameters $\alpha=-5$, $\chi=10$. 
The functional form of $S_{NN}(q)$ in 
Fig.~\ref{Fig:4}(a) indicates a fluid 
structure of the colloidal suspension 
with an increasingly pronounced neighbor 
shell structure upon increasing the 
packing fraction $\ol{\Phi}_{\text{b}}$.
In parallel, the form of $S_{ZZ}(q)$ in 
Fig.~\ref{Fig:4}(b) indicates a spatially 
alternating arrangement of oppositely 
charged colloidal macroions with a 
periodicity of approximately twice the 
nearest-neighbor distance. 

A hypothetical macrophase separation 
of charges would lead to a peak of $S_{ZZ}(q)$ 
at $q=0$, which does obviously not occur here.
From the position of the major peaks in 
Figs.~\ref{Fig:4}(a) and \ref{Fig:4}(b) one 
obtains the mean nearest-neighbor distance 
$\lambda_{NN}$ between two colloidal macroions 
and the periodicity $\lambda_{ZZ}$ of the
charge distribution, respectively; the 
values are displayed in Table~\ref{Tab:1}.
As is expected from the nature of the hard 
core repulsion between two colloidal macroions, 
the nearest-neighbor distance $\lambda_{NN}$ 
is never smaller than the close-contact 
distance $2R_0$. Upon increasing the packing 
fraction $\ol{\Phi}_{\text{b}}$ both $\lambda_{NN}$ 
and $\lambda_{ZZ}$ decrease, but the relation 
$\lambda_{ZZ}\gtrsim2\lambda_{NN}$ holds for 
all cases. Hence the charge distribution 
oscillates on a length scale $\lambda_{ZZ}$
which is approximately twice the nearest-neighbor 
distance $\lambda_{NN}$. This is again showing 
the {\sl alternating charge structure} within the 
suspension of charge-regulated spherical macroions. 
The widths of the major peaks in 
Figs.~\ref{Fig:4}(a) and \ref{Fig:4}(b) 
yield respectively the decay lengths 
(correlation lengths) $\xi_{NN}$ and $\xi_{ZZ}$ 
of the structural features, which are of the 
order of the radius $R_0$ of the colloidal 
macroions, as can be observed in Table~\ref{Tab:1}.
Finally, the coordination number $N_1$, i.e., 
the number of colloidal macroions in the 
nearest-neighbor shell can be calculated 
from $S_{NN}(q)$; the corresponding values 
are displayed in Table~\ref{Tab:1}. 

\begin{figure*}[!t]
\begin{center}
\includegraphics[width=17.5cm]{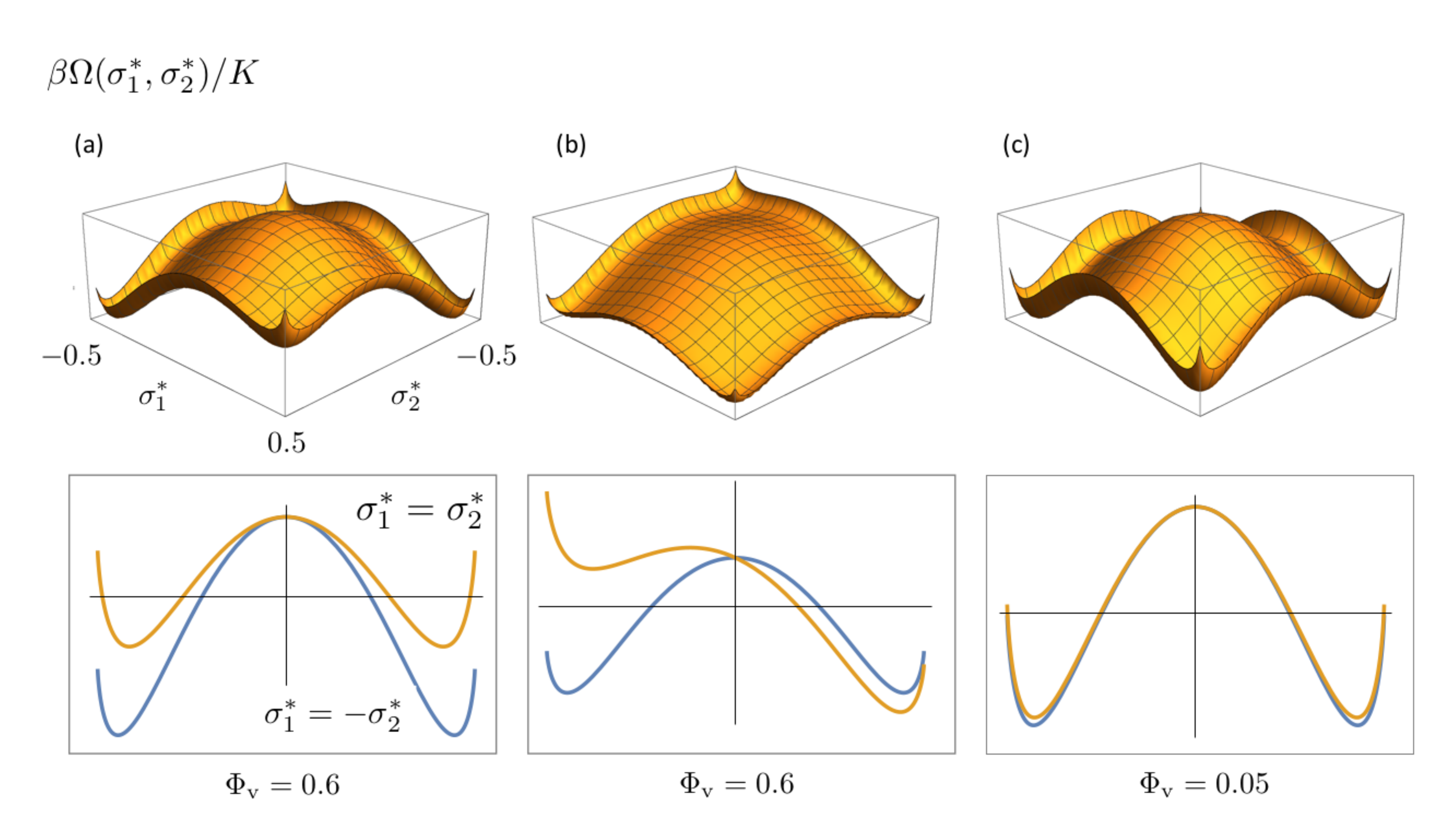}
\end{center}
\caption{Top panel shows the variation of the grand 
         potential $\beta\Omega\left(\sigma_1^*,\sigma_2^*\right)/K$ 
         given by Eq.~\eqref{eq:14} as functions of 
         the surface charge density variables $\sigma_1^*$ 
         and $\sigma_2^*$ for different values of the 
         parameters $\alpha$ and $\Phi_{\text{v}}$. 
         Bottom panel show cuts through the potential 
         surface at $\sigma^*_1=\sigma^*_2$ (in yellow, 
         light shade) and $\sigma^*_1=-\sigma^*_2$ (in 
         blue, dark shade). 
         Panel (a) shows the results for $\alpha=-3$, 
         $\chi=6$ and $\Phi_{\text{v}}=0.6$ which falls 
         inside the phase separation region. Consequently, 
         the asymmetric configuration is the minimum-energy 
         configuration. Panel (b): At the same value of 
         $\Phi_{\text{v}}$, but with $\alpha$ changing from 
         $-3$ to $-2.8$. As it is evident from the curves, the 
         stability has changed towards a configuration with 
         uncoupled cells, thus leaving the phase coexistence 
         region. Panel (c): Again at $\alpha=-3$ and $\chi=6$ 
         but going down in the particle-cell volume ratio to 
         $\Phi_{\text{v}}=0.05$. As one can see, the grand 
         potential of the symmetric configuration approaches 
         that of the asymmetric configuration. For all the plots, 
         $R_0=10\,\mathrm{nm}$, $\kappa=0.1\,\mathrm{nm}^{-1}$, 
         $\ell_B = 0.7\,\mathrm{nm}$ and $K=20$ are used.}   
\label{Fig:5}
\end{figure*}

\section{Binary cell model\label{sec:4}} 

The DFT calculations presented so far 
work the best for dilute suspension 
of charged colloidal macroions. While 
it can also be expected to work well 
for an aggregating system of weakly 
charged macroions, a dense suspension 
of strongly charged macroions surely 
needs to be treated differently. 
Accordingly, in this section we present 
a variation on the standard cell model 
that we refer to as the \textsl{binary 
cell-model} (BCM) that can be invoked 
to describe a dense suspension of 
identical charge-regulated macroions irrespective 
of their surface charge densities and 
is consequently valid for symmetric 
as well as asymmetric charge configurations, 
see Fig.~\ref{Fig:1}. Contrary to the 
standard cell model as well as the cell 
model in binary colloidal mixtures \cite{Tor08}, 
the building block of our variety of the 
cell model are \textsl{two adjacent cells} 
of radius $R$ each of which encloses a 
charged particle of radius $R_0$ (see 
Fig.~\ref{Fig:1}(b)). The macroions 
as well as the wrapping cells are 
considered to be fixed in space.
Such a model inherently assumes 
a certain infinite range crystalline 
order and the validity of such assumption 
clearly improves the closer the system 
is to a high density crystalline state.

The model medium built from this 
elementary cell construct thus has 
a particle-cell volume ratio 
$\Phi_{\text{v}}=(R_0/R)^3$ which 
is related to the bulk packing 
fraction $\ol\Phi_{\text{b}}$ 
defined earlier via the relation 
$\Phi_{\text{v}}=\ol\Phi_{\text{b}}/\Phi_{\text{cp}}$ 
where $\Phi_{\text{cp}}\approx 0.74$ 
is the packing fraction corresponding 
to the {face-centered cubic or} 
hexagonal close-packed arrangement 
of the cells. Clearly, {$\Phi_{\text{v}}$} 
is inversely proportional to the cube 
of the inter-macroion separation. 
The charges on the macroion surfaces 
are again regulated according to 
the mechanism introduced in 
Sec.~\ref{sec:2} (see Eq.~\eqref{eq:1}).

The following considerations are 
based on a Poisson-Boltzmann (PB) 
theory of the binary cell system. 
The grand potential corresponding 
to a \textsl{single} cell, in units 
of the thermal energy $\beta=1/k_BT$, 
can be written as
\begin{align}
 \beta\widehat\Omega\left(\eta,E_R\right)=
 \beta\widehat\Omega^{\text{el}}\left(\eta,E_R\right)+K\beta\widehat\Omega^{\text{s}}\left(\eta\right),
 \label{eq:9}
\end{align}
where $\beta\widehat\Omega^{\text{s}}\left(\eta\right)$ 
is the energy contribution stemming 
from the chemical processes driving 
the (de)protonation reaction (as 
defined in Eq.~\eqref{eq:1}) at the 
macroion surface and
\begin{align}
 \frac{\beta\widehat\Omega^{\text{el}}\left(\eta,E_R\right)}{4\pi R_0^2}=&\frac{-\varepsilon}{\beta e^2}
 \int\limits_{R_0}^{R}dr\frac{r^2}{R_0^2}\left[\kappa^2\cosh\left(\phi\left(r\right)\right)
 +\frac{1}{2}\left(\phi'\left(r\right)\right)^2\right]\nonumber\\
 &+\frac{\phi\left(R_0\right)}{a^2}\left(\eta-\frac{1}{2}\right)-\frac{\varepsilon}{e}\frac{R^2}{R_0^2}E_R\phi\left(R\right)
 \label{eq:10}
\end{align}
with the permittivity 
$\varepsilon=\varepsilon_r\varepsilon_0$
of the embedding medium is the 
electrostatic part of the grand 
potential expressed per unit 
surface area of the macroion 
\cite{Mar21}. Herein $\phi\left(r\right)$ 
is the dimensionless electrostatic 
potential expressed in the units of 
$\beta e$ which fulfills the
PB equation in spherical symmetry,
\begin{align*} 
 \frac{1}{r^2}\left(r^2\phi'(r)\right)' = \kappa^2 \sinh(\phi(r)),
\end{align*}
subject to the boundary condition of a 
charge density $\sigma$ at the macroion 
surface, i.e., at $r=R_0$,
\begin{align}
 \phi'(R_0) = - \frac{\beta e^2}{\varepsilon a^2}\left(\eta-\frac{1}{2}\right)=-\frac{\beta e\sigma}{\varepsilon},
 \label{eq:11}
\end{align}
and of a given radial component $E_R$ 
of the electric field at the cell 
boundary, i.e., at $r=R$,
\begin{align}
 \phi'(R) = - \beta e E_R.
 \label{eq:12}
\end{align}
For {\sl two} coupled cells the grand 
potential of the binary cell system 
can then be written as 
\begin{align}
\beta\Omega\left(\eta_1,\eta_2,E_R\right)=
\beta\widehat\Omega\left(\eta_1,E_R\right)
+\beta\widehat\Omega\left(\eta_2,-E_R\right),
\label{eq:13}
\end{align}
where the subscripts ``1'' and ``2'' are 
used to indicate the two adjacent cells 
in the binary cell model. The energy 
contribution $\beta\widehat\Omega$ for the 
second cell is identical to the first one 
and is given by Eq.~\eqref{eq:9}, albeit  
with the electric field $E_R$ replaced by 
$-E_R$ as the centers of the two adjacent 
cells imply a different unit normal at 
the boundary, see Fig.~\ref{Fig:1}. 

\subsection{Debye-H\"uckel case}

First, we consider the Debye-H\"uckel 
case, i.e., the linearized PB equation, 
which renders the problem analytically 
tractable in part and also straightforwardly 
allows to scan the phase-space spanning 
over the whole ranges of ($\eta_1,\eta_2$) 
or equivalently ($\sigma_1^*,\sigma_2^*$). 

The exact solution of the linear 
electrostatic problem implies a 
nonlinear function 
\begin{align*}
 \displaystyle\beta\Omega(\eta_1,\eta_2)=\min_{E_R}\beta\Omega(\eta_1,\eta_2,E_R)
\end{align*}
of the degrees of protonation 
$\eta_1, \eta_2$, of the binary 
cell model, which subsequently 
needs to be minimized numerically. 
In order to achieve this, one can 
proceed as described in the 
Supplementary Material of \cite{Maj18}, 
where the linearized PB-equation is 
discussed in a planar geometry. 
In the present case we can make 
use of the known solution of the 
linearized PB equation in spherical 
geometry \cite{Bie06}; details are 
described in Appendix \ref{sec:A2}. 
After inserting the equilibrium $E_R$ 
that minimizes the 
$\beta\Omega(\eta_1,\eta_2,E_R)$ 
obtained by expanding Eq.~\eqref{eq:10} 
up to second order in $\phi$ and 
replacing $\phi(r)$ by the known 
solutions, one finally arrives 
at the following expression for 
the grand potential defined in 
Eq.~\eqref{eq:13} as a function 
of $\eta_1$ and $\eta_2$ only:
\begin{align}
\beta\Omega\left(\eta_1,\eta_2\right)\!= 
&\frac{K}{2}\frac{\ell_B}{a^2|\mbox{det~{\bf M}}|}
\left[\gamma\left(\eta_1-\frac{1}{2}\right)^2+\gamma\left(\eta_2-\frac{1}{2}\right)^2\right.\notag\\
&\left.-\frac{\tau^2}{2|\nu|} (\eta_1 - \eta_2)^2\vphantom{\left(\eta_1-\frac{1}{2}\right)^2}\right]
+K\beta\left[\widehat\Omega^{\text{s}}(\eta_1)+\widehat\Omega^{\text{s}}(\eta_2)\right]. 
\label{eq:14}
\end{align}
As before, $K = 4\pi R_0^2/a^2$, and 
$\ell_B$ is the Bjerrum length. The 
factors $\left|\mbox{det~{\bf M}}\right|$, 
$\gamma$, $\tau$ and $\nu$ involve the 
three length scales of the problem: 
the radius of the macroion, the Debye 
length and the cell size; the analytic 
expressions are given in Appendix \ref{sec:A2}.

\begin{figure}[!t]
   \includegraphics[width=8cm]{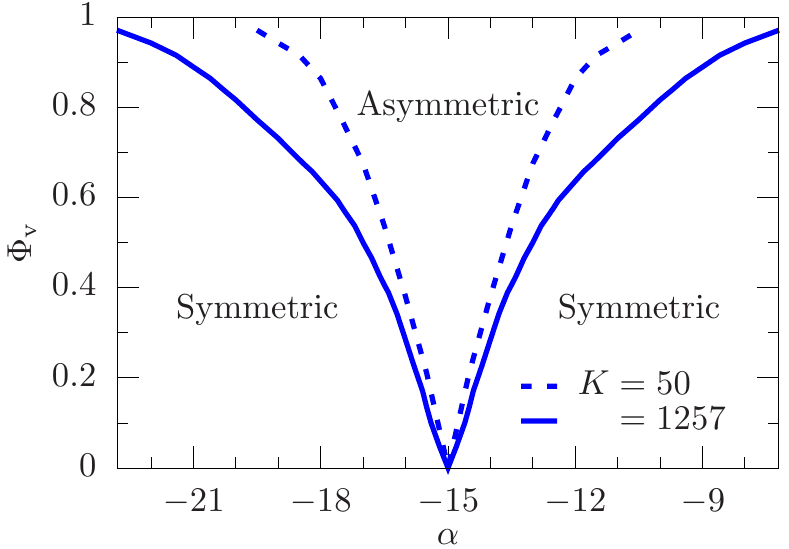}
   \caption{Map showing the charge states of 
   the two macroions as functions of the 
   interaction parameter $\alpha$ and the 
   particle-cell volume fraction $\Phi_{\text{v}}$ 
   for $\chi=30$ and two different $K$-values. 
   The solid line corresponds to $K=1257$ 
   whereas the dashed line refers to $K=50$. 
   In each case, one observes a conical shaped 
   region centered around $\alpha=-\chi/2=-15$ 
   with stretched opening at the top. Inside 
   this region and on the boundary, the two 
   macroion surfaces are oppositely 
   charged ($\left|\eta_1-\eta_2\right|=1$ or 
   equivalently, $\sigma^*_1=-\sigma^*_2$) 
   whereas outside this region they are 
   identically charged ($\eta_1=\eta_2$ or 
   equivalently, $\sigma_1=\sigma_2$). 
   With increasing $K$-value, this region 
   featuring charge-asymmetry widens.}
   \label{Fig:6}
\end{figure}

In line with the DFT calculation and 
the inherent approximations of 
DH theory, we consider the limit of 
small $K$ only. In this regime the 
binary cell model allows for asymmetric 
charge configurations already at smaller 
values of $(\alpha,\chi)$ parameters 
than within the DFT approach in 
Sec.~\ref{sec:3}. The reason for this 
is a weaker electrostatic coupling 
between two colloidal macroions in 
Sec.~\ref{sec:3}, which is based on 
a superposition approximation of the 
electrostatic interaction, as compared
to the stronger coupling {\sl via} the 
electric field $E_R$ at the common 
boundary between two adjacent cells 
within the binary cell model \cite{Maj14, Maj18-1}. 

The phase coexistence around the 
symmetry axis $-2\alpha=\chi$ is 
brought about by an exchange of 
stability of the minima of the grand 
potential, as illustrated in 
Fig.~\ref{Fig:5}. For a given 
particle-cell volume ratio 
$\Phi_{\text{v}}$, the charge 
state with minimum energy shifts 
from a symmetric to an asymmetric 
one as one moves away from the 
symmetry axis $-2\alpha=\chi$ 
(compare {panels (a) and (b)}
of Fig.~\ref{Fig:5}). Moreover, 
a comparison between panels
(a) and (c) of Fig.~\ref{Fig:5} 
suggests that for given 
$(\alpha,\chi)$, as the 
particle-cell volume ratio 
$\Phi_{\text{v}}$ diminishes, 
the difference in the grand 
potential between the symmetric 
and asymmetric configurations 
gradually diminishes too, leading 
ultimately to a symmetric 
equilibrium state.

It is worth noting that the 
linear theory ceases to be valid 
not only for high $K$-values but 
also for higher values of the 
packing fraction $\ol\Phi_{\text{b}}$ 
(or equivalently, the particle-cell 
volume ratio $\Phi_{\text{v}}$) 
where the following nonlinear PB 
theory needs to be applied. 

\subsection{Full Poisson-Boltzmann case}

Within the nonlinear PB theory, 
the equilibrium $\eta$-values 
(or equivalently $\sigma^*$-values;
see Eq.~\eqref{eq:4}) at the two 
surfaces along with the electric 
field at the cell boundary, $E_R$, 
are obtained via a numerical 
minimization of the grand potential 
following the scheme described in 
the beginning of Sec.~\ref{sec:4}. 
We consider a system of macroions 
dispersed in an aqueous electrolyte 
solution with ionic concentration 
$I=1\,\mathrm{mM}$, relative 
permittivity $\varepsilon_r\approx80$ 
at temperature $T=300\,\mathrm{K}$. 

The resulting variations of the 
degrees of protonations $\eta_1$ 
and $\eta_2$ leading to charge 
densities $\sigma^*_1$ and 
$\sigma^*_2$ at the two macroion 
surfaces are shown in Fig.~\ref{Fig:6} 
as functions of the interaction 
parameter $\alpha$ and the 
particle-cell volume ratio 
$\Phi_{\text{v}}$ for two 
different values of the parameter 
$K\in\left\{50, 1257\right\}$ 
corresponding to 
$a\in\left\{5,1\right\}\mathrm{nm}$ 
and $\chi=30$. Note that unlike in 
the case of the linear DH theory, 
we are not constrained by any upper 
limit for $K$ here. Nevertheless, as 
one can infer from Fig.~\ref{Fig:6}, 
the results are qualitatively 
similar to those obtained within 
the linearized PB theory in the 
previous section and are consistent 
with the findings reported in 
Fig.~\ref{Fig:5}. For any given 
$K$-value, one obtains a 
triangularly-shaped region 
centered around $\alpha=-\chi/2$ 
with stretched opening at the top. 
Inside this region and on the 
boundary, the two macroion surfaces 
are oppositely charged 
($\sigma^*_1\approx-\sigma^*_2\neq0$) 
whereas outside this region they 
are identically charged 
($\sigma^*_1=\sigma^*_2$). In 
accordance with the outcomes of 
the linear theory in Sec.~\ref{sec:3}, 
within these identically charged 
regions, both the surfaces are 
negatively charged for 
$\alpha<\frac{\chi}{2}=-15$ whereas 
for $\alpha>\frac{\chi}{2}=-15$, 
they are positively charged. 

With increasing $K$-value, the 
region with asymmetric charge 
configurations broadens as an 
increase in $K$ implies a higher 
surface charge density, which in 
turn enhances the electrostatic 
attraction between the surfaces 
at the origin of the observed 
symmetry breaking. Although in 
general, for a given $\alpha$-value, 
the asymmetric configuration 
changes to the symmetric one 
with decreasing volume fraction 
(or equivalently, increasing 
separation between the macroions), 
the asymmetric configurations 
observed on the line $\alpha=-\chi/2=-15$ 
are very stable and persist down to 
$\Phi_{\text{v}}\approx 10^{-3}$
or lower. The specific choice of the 
parameter $\chi=30$ is motivated by 
the observation that complete symmetry 
breaking, i.e., $\sigma^*_1\approx-\sigma^*_2\neq0$, 
occurs above a critical value 
$\chi=\chi_c\approx25$ for $K=1257$. 
It is important to note that this 
$\chi_c$-value depends not only on $K$ 
but can be different within the 
linear and nonlinear PB theories. 
However, for any $\chi>\chi_c$, one 
observes the same qualitative features, 
i.e., transitions from symmetric to 
symmetry-broken states as functions 
of $\alpha$ and $\Phi_{\text{v}}$ 
within both the theories.

\section{Conclusions\label{sec:5}}

In summary, we have described a 
charge regulation based mechanism 
that allows for pH-dependent phase 
separation in macroion solutions. 
A complex interplay of the different 
chemical interactions driving the 
charge-regulation of individual 
macroion as well as electrostatic 
interaction between them, leads 
to symmetry-broken charge states 
of the macroions, thereby leading 
to aggregation. A density functional 
theory based approach, applicable for 
dilute suspension of macroions, fully 
accounting for the translational  
entropy of the macroions, indeed 
provides evidence of 
electrostatically-driven macroion 
phase separation. A binary cell 
model Poisson-Boltzmann approach, 
applicable in the high density limit, 
confirms the presence of electrostatic 
attraction,  essential for the 
observed phase separation, stemming 
from a transition {to asymmetrically 
charged states} of nearest neighbor 
macroion pairs.

An interesting aside to our 
calculation is the way it naturally 
ties together the well-studied 
complex coacervation between 
chemically oppositely charged 
macroions and the much less studied 
simple coacervation of chemically 
identical macroions in bathing 
electrolyte solutions, by introducing 
the actual chemical model of charging, 
as opposed to \textsl{a priori} chosen 
values of the surface charge (potential). 
We propose this \textsl{charge 
symmetry-broken complex coacervation} 
between chemically identical macroions 
as a bridge between the two different 
types of coacervations or indeed 
liquid-liquid phase separations.

A systematic study based on 
the two contrasting approaches, 
the DFT and the BCM within the 
same charge regulation model,  
therefore ensures the robustness 
of our results and allows us to 
conclude that the pH-dependent 
liquid-liquid phase separation 
in macroion solutions is a rule 
rather than exception even in 
the case of chemically identical 
macroions. Further indications 
for the robustness and generality 
of the described results are 
qualitative similarities with 
approaches based on alternative 
computational methodologies, such 
as the phenomenological Flory-like 
electrostatics \cite{Ara20} and 
the collective mean-field 
description \cite{Avn20}. It 
still remains to be seen which 
are the absolutely essential 
ingredients of the macroion 
surface charge regulation 
promoting this liquid-liquid 
phase separation.

Finally, recent advances in the 
simulations of acid-base equilibria 
in systems coupled to a reservoir 
with a fixed pH, based either on a 
hybrid Monte Carlo method to resolve 
the charges of individual surface 
groups \cite{Cur20}, on the 
grand-reaction method for 
coarse-grained simulations of 
acid-base equilibria with a fixed 
pH reservoir and salt concentration 
\cite{Lan20}, or simulating the pH 
effects with classical coarse-grained 
molecular dynamics simulations 
\cite{Gru20}, could in principle 
provide a proper background to 
different analytical approaches 
and hopefully elucidate the reality 
of the predicted phenomena. The 
comparison with experiments, in 
which other types of interactions 
and the notoriously difficult 
solvent effects come into play, 
poses another challenge that 
will have to be faced in the future.

\begin{acknowledgments}

R.P. would like to acknowledge the 
support of the 1000-Talents Program 
of the Chinese Foreign Experts Bureau. 
He would also like to thank Y.\ Avni 
for her valuable comments. 

\end{acknowledgments} 

\appendix

\section{Density functional theory\label{sec:A1}}

\subsection{Bulk packing fraction}

Upon expanding the term $1-\exp(-\beta U_\text{el})$ 
in Eq.~\eqref{eq:5} of the main text in powers of 
$\beta U_\text{el}$ and using Eq.~\eqref{eq:3} of 
the main text one can perform both 
integrations over $\mathcal{V}$ in Eq.~\eqref{eq:5} 
to obtain
\begin{align}
   \beta F^\text{ex}_\text{el}[n_b]
   = \frac{3V}{4\pi R_0^3} \sum_{k=1}^\infty B_k
   \left(\int\limits_0^1d\eta\,\sigma^*(\eta)^k\Phi_{\text{b}}(\eta)\right)^2
   \label{eq:1_SI}
\end{align}
with the system volume $V = \left|\mathcal{V}\right|$, 
the constants
\begin{align*}
   B_k:=\frac{3}{2}\frac{(-1)^{k-1}k^{k-3}}{k!(\kappa R_0)^3}
   \Gamma(3-k,2k\kappa R_0) 
   \left(K^2\frac{\kappa\ell_B}{1+\kappa R_0}\exp(\kappa R_0)\right)^k,
\end{align*}
where $\Gamma(\nu,z)$ denotes the incomplete 
$\Gamma$-function \cite{Gra80}, and the bulk 
packing fraction profile
\begin{align}
   \Phi_{\text{b}}(\eta) = \frac{4\pi}{3}R_0^3n_{\text{b}}(\eta)
   \label{eq:2_SI}
\end{align}
with $\ol{\Phi}_{\text{b}} = \int\limits_0^1\d\eta\,\Phi_{\text{b}}(\eta)$. 
Using Eqs.~\eqref{eq:1_SI} and \eqref{eq:2_SI} in 
Eq.~\eqref{eq:2} of the main text one obtains 
a scaled density functional in terms of the bulk 
packing fraction profiles $\Phi_{\text{b}}$:
\begin{align}
   \Omega^*_{\text{b}}[\Phi_{\text{b}}]
   :=&\beta\Omega[n_{\text{b}}]\frac{4\pi R_0^3}{3V}
   \notag\\
   =&\int\limits_0^1d\eta\,
   \Phi_{\text{b}}(\eta)\left(\ln(\Phi_{\text{b}}(\eta)) +
   K\beta\widehat\Omega^{\text{s}}\left(\eta\right)\right)\notag\\
   &+\ol{\Phi}_{\text{b}}\left(-1 - \mu^* + h_\text{PY}(\ol{\Phi}_{\text{b}})\right)\notag\\
   &+\sum_{k=1}^\infty B_k\left(\int\limits_0^1d\eta\,\sigma^*(\eta)^k
   \Phi_{\text{b}}(\eta)\right)^2
   \label{eq:3_SI}
\end{align}
with the scaled chemical potential
\begin{align*}
   \mu^* := \ln\left(\frac{4\pi}{3}R_0^3\,\zeta\right)
\end{align*}
and
\begin{align*}
   h_\text{PY}(\ol{\Phi}_{\text{b}}) =
   -\ln\left(1-\ol{\Phi}_{\text{b}}\right) +
   \frac{6\ol{\Phi}_{\text{b}} - 2\ol{\Phi}_{\text{b}}^2}{2\left(1-\ol{\Phi}_{\text{b}}\right)^2}.
\end{align*}
The equilibrium bulk packing fraction profile $\Phi_{\text{b}}^{\text{eq}}$ 
minimizes the scaled density functional $\Omega^*_{\text{b}}$ in 
Eq.~\eqref{eq:3_SI}. Hence it solves the Euler-Lagrange 
equation \cite{Eva79}
\begin{align}
   0
   =&\left.\frac{\delta\Omega^*_{\text{b}}}{\delta\Phi_{\text{b}}(\eta)}\right|_{\Phi_{\text{b}}^\text{eq}}
   \notag\\
   =&\ln\left(\Phi_{\text{b}}^\text{eq}(\eta)\right)\!+\!K\beta\widehat\Omega^{\text{s}}
   \left(\eta\right) - \mu^*\!+\!
   h_\text{PY}(\ol{\Phi}_{\text{b}}^\text{eq})\!+\!\ol{\Phi}_{\text{b}}^\text{eq}
   h_\text{PY}'(\ol{\Phi}_{\text{b}}^\text{eq})\nonumber\\
   &+ 2\sum_{k=1}^\infty B_k\int\limits_0^1\!\!\d\eta'\,\sigma^*(\eta')^k
   \Phi_{\text{b}}^\text{eq}(\eta')\ \sigma^*(\eta)^k.
   \label{eq:4_SI}
\end{align}
This expression can be rewritten in the form
\begin{align*}
   \Phi_{\text{b}}^\text{eq}(\eta) =
   &\exp\bigg(\mu^* - K\beta\widehat\Omega^{\text{s}}\left(\eta\right) - h_\text{PY}(\ol{\Phi}_{\text{b}}^\text{eq})
           - \ol{\Phi}_{\text{b}}^\text{eq}h_\text{PY}'(\ol{\Phi}_{\text{b}}^\text{eq})\nonumber\\
   &-2\sum_{k=1}^\infty B_k\ol{\Phi}_{\text{b}}^\text{eq}\av{\sigma^{*k}}\,
   \sigma^*(\eta)^k\bigg)
\end{align*}
with the $k$-th moment of the surface charge distribution
\begin{align*}
   \av{\sigma^{*k}} := \frac{1}{\ol{\Phi}_{\text{b}}^\text{eq}}
   \int\limits_0^1d\eta\,\sigma^*(\eta)^k\Phi_{\text{b}}^\text{eq}(\eta).
\end{align*}

\subsection{Partial structure factor}

The partial structure factor of the system under consideration can be
written in the form
\begin{align*}
   S(q,\eta,\eta')
   = \frac{1}{\ol{n}_{\text{b}}}\widetilde{G}(q,\eta,\eta')
   =: \frac{\sqrt{\Phi_{\text{b}}(\eta)\Phi_{\text{b}}(\eta')}}{\ol{\Phi}_{\text{b}}}
       \mathcal{G}(q,\eta,\eta'),
\end{align*}
where $\widetilde{G}(q,\eta,\eta')$ is the 
three-dimensional Fourier transform of the bulk 
correlation function $G(r,\eta,\eta')$ of the 
number densities of colloidal spheres with 
degrees of protonation $\eta$ and $\eta'$ at 
a distance $r$. The auxiliary function 
$\mathcal{G}(q,\eta,\eta')$ fulfills the
Ornstein-Zernike equation
\begin{align*}
   \mathcal{G}(q,\eta,\eta') =
   \int\limits_0^1\!\!\d\eta''\,\mathcal{C}(q,\eta,\eta'')
   \mathcal{G}(q,\eta'',\eta') + \delta(\eta-\eta')
\end{align*}
with $\mathcal{C}(q,\eta,\eta')=\!\sqrt{n_{\text{b}}(\eta)n_{\text{b}}(\eta')}\,
\widetilde{c}(q,\eta,\eta')$, where $\widetilde{c}(q,\eta,\eta')$, 
$q=\left|\mathbf{q}\right|$, is the Fourier transform of the bulk direct 
correlation function
\begin{align*}
   c(\mathbf{r},\eta,\eta') = c_\text{PY}(\mathbf{r},\ol{\Phi}_{\text{b}})
   - \frac{\delta^2\beta F^\text{ex}_\text{el}}{\delta n(\mathbf{r},\eta)
   \delta n(\mathbf{0},\eta')}\bigg|_{n_{\text{b}}}
\end{align*}
composed of the Percus-Yevick hard-core 
contribution $c_\text{PY}$ (see Ref.~\cite{Han86}) 
and the electrostatic contribution obtained 
from Eq.~\eqref{eq:5} of the main text (see Ref.~\cite{Eva79}).

\section{Debye-H\"uckel theory for binary cell model\label{sec:A2}}

In this section we show how 
Eq.~\eqref{eq:14} of the main text can 
be derived. First, the linearized 
PB-term integral
\begin{align*}
\int_{R_0}^{R} dr r^2 \left[(\phi'(r))^2 + \kappa^2 (\phi(r))^2\right],
\end{align*}
can be simplified by partial integration, since one has
\begin{align*}
\int_{R_0}^R dr (r^2 \phi'(r))\phi'(r)  =& \left.r^2 \phi'(r)\phi(r)\right|_{R_0}^{R}\nonumber\\ 
&- \int_{R_0}^R dr \partial_r (r^2 \partial_r \phi(r)) \phi(r), 
\end{align*}
whereby the latter term can be 
transformed via the linearized 
PB equation and hence cancels 
out against the term in the 
original integral. Putting all 
constants and the boundary 
conditions in, one is left 
with the expression
\begin{align*}
\frac{\beta \widehat{\Omega}^{\text{el}}}{4\pi R_0^2} \equiv -\frac{\varepsilon}{2\beta e^2} \left(\frac{R}{R_0}\right)^2E_R\phi(R) + \frac{\sigma^*}{2a^2}\phi(R_0). 
\end{align*}
Coupling two cells, respecting electroneutrality, 
\begin{align*}
{\Omega}^{\text{el}} = \widehat{\Omega}^{\text{el}}_1\left([\phi_1],E_R,\sigma_1^*\right) + \widehat{\Omega}^{\text{el}}_2\left([\phi_2],-E_R,\sigma_2^*\right)
\end{align*}
one has the expression 
\begin{align}
\frac{\beta{\Omega}^{\text{el}}}{4\pi R_0^2}=&
-\frac{\varepsilon}{2\beta e^2} \left(\frac{R}{R_0}\right)^2E_R(\phi_1(R) - \phi_2(R))\nonumber\\ 
&+ \frac{1}{2a^2}(\sigma_1^*\phi_1(R_0) + \sigma_2^*\phi_2(R_0)). 
\label{eq:7_SI}
\end{align}
The boundary conditions at the cell particles 
and surfaces follow from the known exact solution 
of the electrostatic potential of a single cell 
with a solute of radius $R_0$, embedded in a 
spherical cell of radius $R$ which contains a 
salt solution, as given in \cite{Bie06}:
\begin{align*}
\phi(r)  =& \frac{R_0 \phi(R_0)}{r} \cosh(\kappa(r - R_0))\nonumber\\
&+ \frac{R \phi(R) - R_0 \phi(R_0) \cosh(\kappa(R-R_0))}{\sinh(\kappa(R-r_0))} \sinh(\kappa(r-R_0)).
\end{align*} 
The boundary values $\phi(R_0)$ and $\phi(R)$ 
thus follow from the derivatives of $\phi(r)$ 
at $r = R_0, R$. 

\begin{figure}
\begin{center}
\includegraphics[width=8cm]{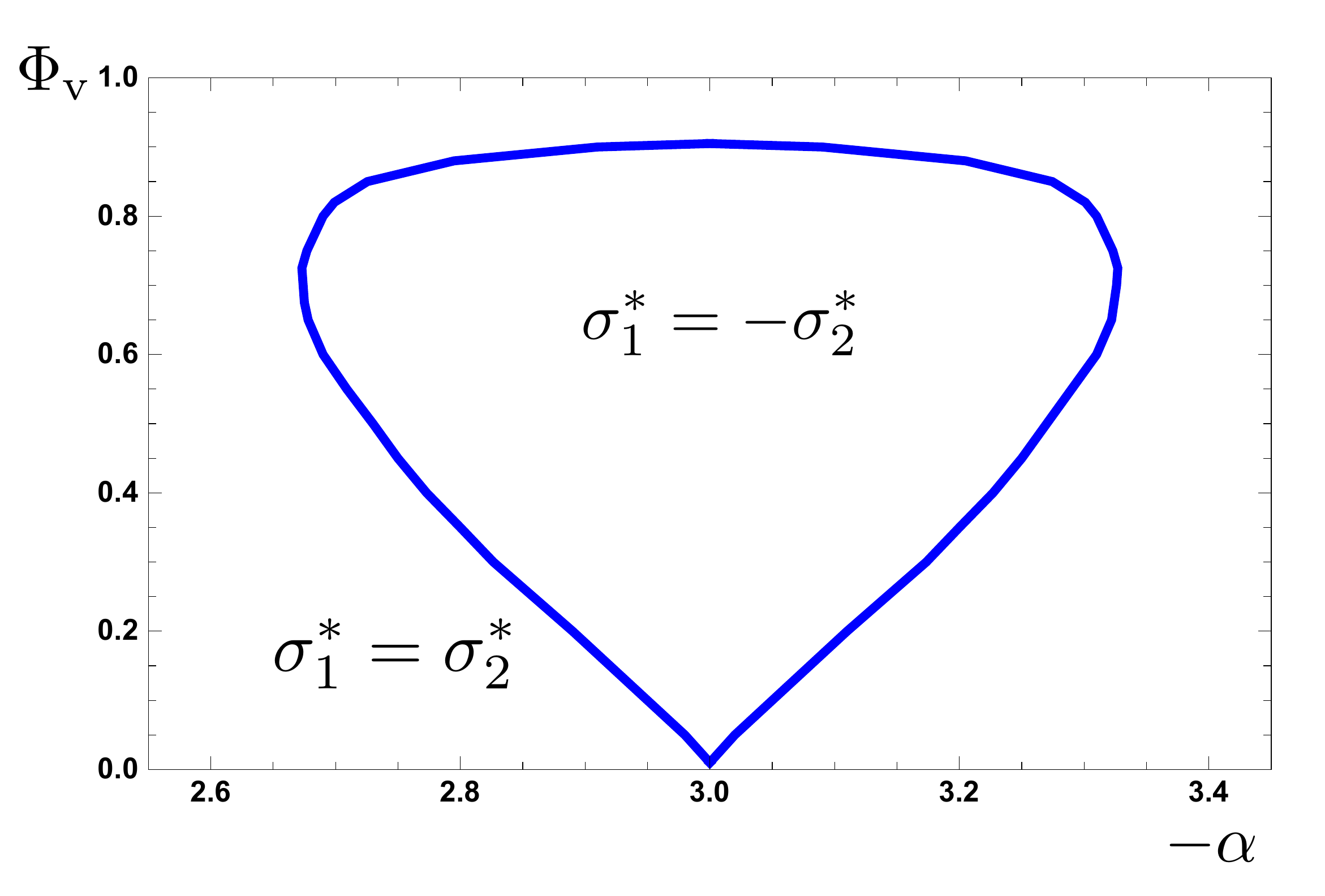}
\end{center}
\caption{Two-phase surface-charge density 
coexistence region of around the symmetry 
axis with $\alpha = -3$, at $\chi = 6$ for 
$K=20$.} 
\label{Fig:7}
\end{figure}

In the linear theory we discuss, denoting 
the vector ${\bf \phi'} = (\phi'(R_0), \phi'(R))$ 
of derivatives, one needs to invert the 
matrix equation 
${\bf \phi'} = {\bf M} \cdot {\bf \phi}$ which 
can be computed from the derivative of the 
solution. One finds
\begin{align*} 
\phi_i(R_0) = \frac{1}{\mbox{det}\, {\bf M}} [\gamma \phi'_i(R_0) - \tau \phi'_i(R)],   
\end{align*}
and
\begin{align*}
\phi_i(R) = \frac{1}{\mbox{det}\, {\bf M}} [-\xi \phi'_i(R_0) + \nu \phi'_i(R)],   
\end{align*}
where the boundary values for the two cells 
are given by Eqs.~\eqref{eq:11} and 
\eqref{eq:12} of the main text. The parameters 
in the equations are functions of the three 
characteristic lengths in the system (solute 
size, screening length, cell radius):
\begin{align*}
\nu & =   - \left[\frac{1}{R_0} + \frac{\kappa C(R)}{S(R)}\right]\,,\,\,  \tau = \frac{R}{R_0}\frac{\kappa}{S(R)}\,,\,\, \nonumber \\
\xi & =   - \frac{R_0}{R}\frac{\kappa}{S(R)}\,,\,\, \gamma = - \frac{1}{R} + \frac{\kappa C(R)}{S(R)}, 
\end{align*}
where $C(R) = \cosh(\kappa(R-R_0))$ and 
$S(R) = \sinh(\kappa(R - R_0))$. Further,
\begin{align*}
\mbox{det}\, {\bf M} = \nu \gamma - \tau \xi\, .
\end{align*}
One has $\nu < 0$, $\tau > 0 $, $\xi < 0$, $\gamma > 0$ 
and $\mbox{det}\, {\bf M} < 0$. 
After computation of the boundary conditions 
at the macroion radius $R_0$ and the binary 
cell radius $R$, the resulting expression 
Eq.~\eqref{eq:7_SI} needs to be minimized 
with respect to $E_R$, which is found to 
behave as $E_R\sim\sigma_1^*-\sigma_2^*$. 
Finally, collection of terms leads to 
Eq.~\eqref{eq:14} of the main text. 

Within DH-theory, the surface-charge density 
coexistence curve has a balloon-like shape 
(see Fig.~\ref{Fig:7}). The upper part of the 
surface-charge density coexistence curve 
widens for increasing $\chi$. The location 
of the upper critical point of the phase 
coexistence curve moves from 
$(\chi,\Phi_{\text v}) = (3.6,0.79)$ to 
$(\chi,\Phi_{\text v}) = (16, 0.97)$, 
which for $K=20$ covers the interval 
in which surface-charge density coexistence 
exists in the DH-limit.

\end{document}